\documentclass[aps,prx,twocolumn,groupedaddress,superscriptaddress]{revtex4-2}
\usepackage{graphicx}
\usepackage{physics}
\usepackage{amsfonts}
\usepackage{float}
\usepackage{hyperref}
\usepackage[caption=false]{subfig}
\usepackage[all]{hypcap}
\usepackage{xcolor}
\usepackage{algorithm}
\usepackage{algpseudocode}
\usepackage[normalem]{ulem}
\usepackage{tikz}

\definecolor{myblue}{HTML}{73A5FD}
\definecolor{myred}{HTML}{D91B60}
\definecolor{myyellow}{HTML}{FFC008}

\newcommand{\Ac}{\mathcal A}

\newcommand{\Hc}{\mathcal H}

\newcommand{\Xc}{\mathcal X}

\renewcommand{\a}{\alpha}

\renewcommand{\l}{\lambda}
\newcommand{\s}{\sigma}

\DeclareMathOperator*{\argmin}{arg \, min}

\begin{document}


\title{Variational adiabatic transport of tensor networks}


\author{Hyeongjin Kim}
\email[]{hkim12@bu.edu}
\affiliation{Department of Physics, Boston University, Boston, Massachusetts 02215, USA}
\affiliation{Center for Computational Quantum Physics, Flatiron Institute, New York, New York 10010, USA}

\author{Matthew Fishman}
\affiliation{Center for Computational Quantum Physics, Flatiron Institute, New York, New York 10010, USA}

\author{Dries Sels}
\affiliation{Center for Computational Quantum Physics, Flatiron Institute, New York, New York 10010, USA}
\affiliation{Department of Physics, New York University, New York, New York 10003, USA}


\date{\today}

\begin{abstract} We discuss a tensor network method for constructing the adiabatic gauge potential -- the generator of adiabatic transformations -- as a matrix product operator, which allows us to adiabatically transport matrix product states. Adiabatic evolution of tensor networks offers a wide range of applications, of which two are explored in this paper: improving tensor network optimization and scanning phase diagrams. By efficiently transporting eigenstates to quantum criticality and performing intermediary density matrix renormalization group (DMRG) optimizations along the way, we demonstrate that we can compute ground and low-lying excited states faster and more reliably than a standard DMRG method at or near quantum criticality. We demonstrate a simple automated step size adjustment and detection of the critical point based on the norm of the adiabatic gauge potential. Remarkably, we are able to reliably transport states through the critical point of the models we study. 
\end{abstract}


\maketitle

\section{Introduction}

In the past few decades, the density matrix renormalization group (DMRG) \cite{white1992density} has established itself as one of most prominent numerical algorithms for simulating one-dimensional quantum lattice systems (see Refs. \cite{schollwock2011density,verstraete2008matrix,mcculloch2007density} for reviews). It is commonly realized as a variational tensor network method that uses a matrix product state (MPS) as an ansatz for quantum wave functions and has been extensively used to compute low energy states and dynamics of various low-dimensional quantum systems. Determining excited states provides important physical information of the system: for example, excited energy spectra can provide information on quantum phase transitions and thus help determine quantum critical points (the critical points) \cite{wang2018critical,yang2022quantum} or can be used to construct the low temperature spectral function. While ground state methods such as DMRG are well established, excited state methods are still in development. Even so, there have been numerous extensions of DMRG that find excited states accurately and efficiently. One natural extension is to compute the lowest energy states of the Hamiltonian in multiple symmetry sectors, provided that they exist \cite{white1993density,schollwock2011density}. Another, more general, method involves adding a penalty term in the Hamiltonian to penalize, already found, lower energy states~\cite{mcculloch2007density}. There is a plethora of other extensions including but not limited to Refs. \cite{white1993density,ostlund1995thermodynamic,haegeman2012variational,banuls2013mass,hu2015excited,khemani2016obtaining,roberts2017implementation,van2021efficient,li2023tangent,li2023accurate}. However, even with these efforts, computing excited states remains a computationally challenging task, especially near the critical points where the spectrum is dense. 

\begin{figure}[!htp]
\centering
\includegraphics[width=\columnwidth]{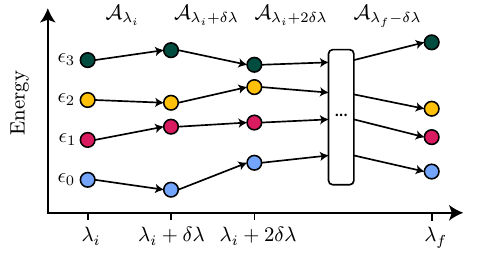}%
\caption[]{Adiabatic evolution of energy eigenstates using the adiabatic gauge potential. As an illustrative example, we begin with four lowest-lying states at $\l_i$. Then, we compute the AGP $\Ac_{\l_i}$ at $\l_i$ to propagate these states to the next parameter point $\l_i + \delta \l$ and continue this procedure iteratively until we reach $\l_f$. In practice, the spacing $\delta \l$ can be automatically adjusted based on the norm of the AGP to add more steps where necessary, for example near quantum critical points, as discussed in Sec. \ref{agp-evo}. Optimization algorithms like DMRG can be used to find more accurate states during the course of the adiabatic evolution.}
\label{fig:propagate}
\end{figure}

Adiabaticity plays a fundamental role in physics with numerous applications in quantum state preparation, quantum computation, and even heat engines in thermodynamics. Specifically, the adiabatic theorem states that a quantum system, prepared in an eigenstate of a Hamiltonian, stays an eigenstate of the Hamiltonian as long as the Hamiltonian is varied slowly enough. One way to formalize this adiabatic transport is through the adiabatic gauge potential (AGP)~\cite{kolodrubetz2017geometry,Berry_2009,adiabatic_population}. In short, the AGP is the generator of adiabatic eigenstate deformations and has been used for various applications, ranging from counterdiabatic driving for optimal control \cite{campo2012assisted,saberi2014adiabatic,sels2017minimizing,claeys2019floquet,vcepaite2023counterdiabatic} to studying quantum chaos of many-body systems \cite{pandey2020adiabatic}.  
Here, we demonstrate that the AGP can be efficiently and reliably represented numerically as a tensor network, specifically a matrix product operator (MPO), and introduce a simple numerical optimization method for obtaining the MPO representation of the AGP. This provides a numerically efficient and general way of performing adiabatic evolution of MPS, with straightforward generalizations to other tensor networks, by evolving with the AGP (in contrast to previous work on adiabatic evolution of tensor networks \cite{Schwarz2012preparing,Schwarz2013preparing,Ge2016rapid,cruz2022preparation,wei2023efficient} based on parent Hamiltonian constructions) and thereby opens up numerous applications in tensor networks. 

 We demonstrate that AGP evolution of MPS can be utilized to significantly improve the reliability and efficiency of computing low energy excited states represented as MPSs at or near criticality, focusing in this work on a particular flavor of excited state DMRG. Our approach involves computing MPS excitations far from the the critical point and adiabatically evolving these states towards the critical point, as shown schematically in Fig. \ref{fig:propagate}, where calculations become significantly more difficult. We show that this method finds a set of low energy eigenstates more reliably, accurately, and efficiently at the critical point than a standard excited state DMRG method. Relatedly, we demonstrate how adiabatic evolution of tensor networks with the AGP can be used to automatically scan the entire phase diagram of quantum many-body systems by evolving low-lying energy eigenstates through different phases. We demonstrate that the norm of the AGP can be used to automatically adjust the step size of the adiabatically evolution, with a higher norm and therefore more steps needed at or near the critical point, giving a way to automatically detect different phases and phase transitions.

\section{\label{agp}Tensor Network Representation of the Adiabatic Gauge Potential}
In this section, we give a brief introduction to the AGP with emphasis on its role as a generator of adiabatic eigenstate deformations and the minimization approach for its construction. While much of this has been discussed in earlier papers \cite{kolodrubetz2017geometry,claeys2019floquet,serbyn22}, it is provided here as context for when we derive a method for constructing the AGP as an MPO. The latter was recently also discussed in detail in Ref.~\cite{serbyn22}. 

\subsection{Background on the adiabatic gauge potential}

Suppose we have a Hamiltonian $\Hc(\lambda)$ with a tunable parameter $\lambda$. The AGP, denoted as $\Ac_\l$, is a generator of adiabatic eigenstate deformations from changing $\l$ via
\begin{equation}\label{eqn:agp}
	i \partial_\l \ket{n(\l)} = \Ac_\l \ket{n(\l)} \, ,
\end{equation}
where $\ket{n(\l)}$ is an energy eigenstate of $\Hc(\l)$ and, in general, the AGP depends on $\lambda$. Using this, we can adiabatically transform the eigenstates of $\Hc(\l)$ to those of $\Hc(\l + \dd{\l})$ with
\begin{equation}\label{eqn:agp-propagate}
	\ket{n(\l +  \dd{\l})} = e^{-i \Ac_\l \dd{\l}} \ket{n(\l)} \, ,
\end{equation}
where $\dd{\l}$ is an infinitesimal perturbation on $\l$.

While the AGP can be numerically computed exactly, it typically requires exact diagonalization, which is computationally infeasible for large systems. To combat this, variational approaches for approximately computing the AGP have been explored in earlier work \cite{saberi2014adiabatic,sugiura2021adiabatic,claeys2019floquet,sels2017minimizing,vcepaite2023counterdiabatic}, which usually involve minimizing the action $S_\l$
\begin{equation}
	\Ac_\l = \argmin \limits_{\Xc} S_\l (\Xc) \, ,
\end{equation}
where 
\begin{equation}\label{eqn:action}
	S_\l(\Xc) = \norm{G_\l(\Xc)}^2 =  \norm{\partial_\l \Hc + i \comm{\Xc}{\Hc}}^2 \, ,
\end{equation}
and $\norm{\cdot}^2$ is the Frobenius norm. To compute $\Ac_\l$, we take the gradient of $S_\l$
\begin{align}
\begin{split}\label{eqn:action_gradient}
    \grad_\Xc{S_\l(\Xc)} &= 2\left(\Xc \Hc^2 + \Hc^2 \Xc - 2 \Hc \Xc \Hc \right) \\
    & \qquad - 2i \left(\partial_\l \Hc \, \Hc - \Hc \, \partial_\l \Hc \right) \, ,
\end{split}
\end{align}
and set it equal to zero. After setting $\grad_\Xc{S_\l(\Xc)} = 0$ in Eq. (\ref{eqn:action_gradient}), we solve for $\Xc$ in the linear equation
\begin{equation}\label{eqn:linear_equation}
    \Xc \Hc^2 + \Hc^2 \Xc - 2\Hc \Xc \Hc = i\left(\partial_\l \Hc \, \Hc - \Hc \, \partial_\l \Hc \right) \, ,
\end{equation}
where we used the fact that $\Hc^\dagger = \Hc$. One method of solving this linear equation is to use a variational ansatz involving the commutator expansion of the form \cite{claeys2019floquet} 
\begin{equation}\label{eqn:agp-comm}
	\Ac^{(\ell)}_\l = i \sum^{\ell}_{k=1} \alpha_k  \underbrace{[\Hc , [\Hc , \ldots [\Hc}_{2k-1}, \, \partial_\l \Hc]]] \, , 
\end{equation}
where the coefficients $\{\a_1, \a_2, \ldots , \a_k\}$ can be obtained from solving Eq. (\ref{eqn:linear_equation}). Notably, the exact AGP is recovered when $\ell \to \infty$. Due to the growing terms in the summation, the calculations quickly becomes inefficient for sufficiently large $\ell$ and the convergence can be slow. Instead, we propose a tensor network method for directly computing the approximate AGP as an MPO whose local support depends on its bond dimension $D$. This allows us to construct the AGP for large system sizes in a well-controlled manner.

\subsection{Adiabatic gauge potential as a matrix product operator}

\begin{figure*}[!htbp]
\centering
\resizebox{1.5\columnwidth}{!}{%
    \begin{tikzpicture}
        $\draw[draw=black,solid,line width=2pt] (1.2, 1.2) -- (2.4, 1.2);$$\draw[draw=black,solid,line width=2pt] (1.2, 1.2) -- (1.2, 2.4);$$\draw[draw=black,solid,line width=2pt] (2.4, 1.2) -- (3.5999999999999996, 1.2);$$\draw[draw=black,solid,line width=2pt] (2.4, 1.2) -- (2.4, 2.4);$$\draw[draw=black,solid,line width=2pt] (3.5999999999999996, 1.2) -- (3.5999999999999996, 2.4);$$\draw[draw=black,solid,line width=2pt] (1.2, 2.4) -- (2.4, 2.4);$$\draw[draw=black,solid,line width=2pt] (1.2, 2.4) -- (1.2, 3.5999999999999996);$$\draw[draw=black,solid,line width=2pt] (2.4, 2.4) -- (3.5999999999999996, 2.4);$$\draw[draw=black,solid,line width=2pt] (2.4, 2.4) -- (2.4, 3.5999999999999996);$$\draw[draw=black,solid,line width=2pt] (3.5999999999999996, 2.4) -- (3.5999999999999996, 3.5999999999999996);$$\draw[draw=black,solid,line width=2pt] (1.2, 3.5999999999999996) -- (2.4, 3.5999999999999996);$$\draw[draw=black,solid,line width=2pt] (2.4, 3.5999999999999996) -- (3.5999999999999996, 3.5999999999999996);$$\draw[draw=black,solid,line width=2pt] (1.2, 1.2) -- (1.2, 0.3999999999999999);$$\draw[draw=black,solid,line width=2pt] (1.2, 1.2) -- (1.2, 2.0);$$\filldraw[draw=black,solid,line width=2pt,fill=myred!90] (1.2, 1.2) circle (0.4) node {};$\draw (1.2, 1.2) node[text=black] {\large $\mathcal{H}_{1}$};$\draw[draw=black,solid,line width=2pt] (2.4, 1.2) -- (2.4, 0.3999999999999999);$$\draw[draw=black,solid,line width=2pt] (2.4, 1.2) -- (2.4, 2.0);$$\filldraw[draw=black,solid,line width=2pt,fill=myred!90] (2.4, 1.2) circle (0.4) node {};$\draw (2.4, 1.2) node[text=black] {\large $\mathcal{H}_{2}$};$\draw[draw=black,solid,line width=2pt] (3.5999999999999996, 1.2) -- (3.5999999999999996, 0.3999999999999999);$$\draw[draw=black,solid,line width=2pt] (3.5999999999999996, 1.2) -- (3.5999999999999996, 2.0);$$\filldraw[draw=black,solid,line width=2pt,fill=myred!90] (3.5999999999999996, 1.2) circle (0.4) node {};$\draw (3.5999999999999996, 1.2) node[text=black] {\large $\mathcal{H}_{3}$};$\draw[draw=black,solid,line width=2pt] (1.2, 2.4) -- (1.2, 1.5999999999999999);$$\draw[draw=black,solid,line width=2pt] (1.2, 2.4) -- (1.2, 3.2);$$\filldraw[draw=black,solid,line width=2pt,fill=myred!90] (1.2, 2.4) circle (0.4) node {};$\draw (1.2, 2.4) node[text=black] {\large $\mathcal{H}_{1}$};$\draw[draw=black,solid,line width=2pt] (2.4, 2.4) -- (2.4, 1.5999999999999999);$$\draw[draw=black,solid,line width=2pt] (2.4, 2.4) -- (2.4, 3.2);$$\filldraw[draw=black,solid,line width=2pt,fill=myred!90] (2.4, 2.4) circle (0.4) node {};$\draw (2.4, 2.4) node[text=black] {\large $\mathcal{H}_{2}$};$\draw[draw=black,solid,line width=2pt] (3.5999999999999996, 2.4) -- (3.5999999999999996, 1.5999999999999999);$$\draw[draw=black,solid,line width=2pt] (3.5999999999999996, 2.4) -- (3.5999999999999996, 3.2);$$\filldraw[draw=black,solid,line width=2pt,fill=myred!90] (3.5999999999999996, 2.4) circle (0.4) node {};$\draw (3.5999999999999996, 2.4) node[text=black] {\large $\mathcal{H}_{3}$};$\draw[draw=black,solid,line width=2pt] (1.2, 3.5999999999999996) -- (1.2, 2.8);$$\draw[draw=black,solid,line width=2pt] (1.2, 3.5999999999999996) -- (1.2, 4.3999999999999995);$$\filldraw[draw=black,solid,line width=2pt,fill=myblue!90] (1.2, 3.5999999999999996) circle (0.4) node {};$\draw (1.2, 3.5999999999999996) node[text=black] {\large $\mathcal{X}_{1}$};$\draw[draw=black,solid,line width=2pt] (2.4, 3.5999999999999996) -- (2.4, 2.8);$$\draw[draw=black,solid,line width=2pt] (2.4, 3.5999999999999996) -- (2.4, 4.3999999999999995);$$\filldraw[draw=black,solid,line width=2pt,fill=myblue!90] (2.4, 3.5999999999999996) circle (0.4) node {};$\draw (2.4, 3.5999999999999996) node[text=black] {\large $\mathcal{X}_{2}$};$\draw[draw=black,solid,line width=2pt] (3.5999999999999996, 3.5999999999999996) -- (3.5999999999999996, 2.8);$$\draw[draw=black,solid,line width=2pt] (3.5999999999999996, 3.5999999999999996) -- (3.5999999999999996, 4.3999999999999995);$$\filldraw[draw=black,solid,line width=2pt,fill=myblue!90] (3.5999999999999996, 3.5999999999999996) circle (0.4) node {};$\draw (3.5999999999999996, 3.5999999999999996) node[text=black] {\large $\mathcal{X}_{3}$};\draw (4.5, 2.4) node[text=black] {\large $+$};$\draw[draw=black,solid,line width=2pt] (5.4, 1.2) -- (6.6, 1.2);$$\draw[draw=black,solid,line width=2pt] (5.4, 1.2) -- (5.4, 2.4);$$\draw[draw=black,solid,line width=2pt] (6.6, 1.2) -- (7.8, 1.2);$$\draw[draw=black,solid,line width=2pt] (6.6, 1.2) -- (6.6, 2.4);$$\draw[draw=black,solid,line width=2pt] (7.8, 1.2) -- (7.8, 2.4);$$\draw[draw=black,solid,line width=2pt] (5.4, 2.4) -- (6.6, 2.4);$$\draw[draw=black,solid,line width=2pt] (5.4, 2.4) -- (5.4, 3.5999999999999996);$$\draw[draw=black,solid,line width=2pt] (6.6, 2.4) -- (7.8, 2.4);$$\draw[draw=black,solid,line width=2pt] (6.6, 2.4) -- (6.6, 3.5999999999999996);$$\draw[draw=black,solid,line width=2pt] (7.8, 2.4) -- (7.8, 3.5999999999999996);$$\draw[draw=black,solid,line width=2pt] (5.4, 3.5999999999999996) -- (6.6, 3.5999999999999996);$$\draw[draw=black,solid,line width=2pt] (6.6, 3.5999999999999996) -- (7.8, 3.5999999999999996);$$\draw[draw=black,solid,line width=2pt] (5.4, 1.2) -- (5.4, 0.3999999999999999);$$\draw[draw=black,solid,line width=2pt] (5.4, 1.2) -- (5.4, 2.0);$$\filldraw[draw=black,solid,line width=2pt,fill=myblue!90] (5.4, 1.2) circle (0.4) node {};$\draw (5.4, 1.2) node[text=black] {\large $\mathcal{X}_{1}$};$\draw[draw=black,solid,line width=2pt] (6.6, 1.2) -- (6.6, 0.3999999999999999);$$\draw[draw=black,solid,line width=2pt] (6.6, 1.2) -- (6.6, 2.0);$$\filldraw[draw=black,solid,line width=2pt,fill=myblue!90] (6.6, 1.2) circle (0.4) node {};$\draw (6.6, 1.2) node[text=black] {\large $\mathcal{X}_{2}$};$\draw[draw=black,solid,line width=2pt] (7.8, 1.2) -- (7.8, 0.3999999999999999);$$\draw[draw=black,solid,line width=2pt] (7.8, 1.2) -- (7.8, 2.0);$$\filldraw[draw=black,solid,line width=2pt,fill=myblue!90] (7.8, 1.2) circle (0.4) node {};$\draw (7.8, 1.2) node[text=black] {\large $\mathcal{X}_{3}$};$\draw[draw=black,solid,line width=2pt] (5.4, 2.4) -- (5.4, 1.5999999999999999);$$\draw[draw=black,solid,line width=2pt] (5.4, 2.4) -- (5.4, 3.2);$$\filldraw[draw=black,solid,line width=2pt,fill=myred!90] (5.4, 2.4) circle (0.4) node {};$\draw (5.4, 2.4) node[text=black] {\large $\mathcal{H}_{1}$};$\draw[draw=black,solid,line width=2pt] (6.6, 2.4) -- (6.6, 1.5999999999999999);$$\draw[draw=black,solid,line width=2pt] (6.6, 2.4) -- (6.6, 3.2);$$\filldraw[draw=black,solid,line width=2pt,fill=myred!90] (6.6, 2.4) circle (0.4) node {};$\draw (6.6, 2.4) node[text=black] {\large $\mathcal{H}_{2}$};$\draw[draw=black,solid,line width=2pt] (7.8, 2.4) -- (7.8, 1.5999999999999999);$$\draw[draw=black,solid,line width=2pt] (7.8, 2.4) -- (7.8, 3.2);$$\filldraw[draw=black,solid,line width=2pt,fill=myred!90] (7.8, 2.4) circle (0.4) node {};$\draw (7.8, 2.4) node[text=black] {\large $\mathcal{H}_{3}$};$\draw[draw=black,solid,line width=2pt] (5.4, 3.5999999999999996) -- (5.4, 2.8);$$\draw[draw=black,solid,line width=2pt] (5.4, 3.5999999999999996) -- (5.4, 4.3999999999999995);$$\filldraw[draw=black,solid,line width=2pt,fill=myred!90] (5.4, 3.5999999999999996) circle (0.4) node {};$\draw (5.4, 3.5999999999999996) node[text=black] {\large $\mathcal{H}_{1}$};$\draw[draw=black,solid,line width=2pt] (6.6, 3.5999999999999996) -- (6.6, 2.8);$$\draw[draw=black,solid,line width=2pt] (6.6, 3.5999999999999996) -- (6.6, 4.3999999999999995);$$\filldraw[draw=black,solid,line width=2pt,fill=myred!90] (6.6, 3.5999999999999996) circle (0.4) node {};$\draw (6.6, 3.5999999999999996) node[text=black] {\large $\mathcal{H}_{2}$};$\draw[draw=black,solid,line width=2pt] (7.8, 3.5999999999999996) -- (7.8, 2.8);$$\draw[draw=black,solid,line width=2pt] (7.8, 3.5999999999999996) -- (7.8, 4.3999999999999995);$$\filldraw[draw=black,solid,line width=2pt,fill=myred!90] (7.8, 3.5999999999999996) circle (0.4) node {};$\draw (7.8, 3.5999999999999996) node[text=black] {\large $\mathcal{H}_{3}$};\draw (8.7, 2.4) node[text=black] {\large $-2$};$\draw[draw=black,solid,line width=2pt] (9.6, 1.2) -- (10.8, 1.2);$$\draw[draw=black,solid,line width=2pt] (9.6, 1.2) -- (9.6, 2.4);$$\draw[draw=black,solid,line width=2pt] (10.8, 1.2) -- (12.0, 1.2);$$\draw[draw=black,solid,line width=2pt] (10.8, 1.2) -- (10.8, 2.4);$$\draw[draw=black,solid,line width=2pt] (12.0, 1.2) -- (12.0, 2.4);$$\draw[draw=black,solid,line width=2pt] (9.6, 2.4) -- (10.8, 2.4);$$\draw[draw=black,solid,line width=2pt] (9.6, 2.4) -- (9.6, 3.5999999999999996);$$\draw[draw=black,solid,line width=2pt] (10.8, 2.4) -- (12.0, 2.4);$$\draw[draw=black,solid,line width=2pt] (10.8, 2.4) -- (10.8, 3.5999999999999996);$$\draw[draw=black,solid,line width=2pt] (12.0, 2.4) -- (12.0, 3.5999999999999996);$$\draw[draw=black,solid,line width=2pt] (9.6, 3.5999999999999996) -- (10.8, 3.5999999999999996);$$\draw[draw=black,solid,line width=2pt] (10.8, 3.5999999999999996) -- (12.0, 3.5999999999999996);$$\draw[draw=black,solid,line width=2pt] (9.6, 1.2) -- (9.6, 0.3999999999999999);$$\draw[draw=black,solid,line width=2pt] (9.6, 1.2) -- (9.6, 2.0);$$\filldraw[draw=black,solid,line width=2pt,fill=myred!90] (9.6, 1.2) circle (0.4) node {};$\draw (9.6, 1.2) node[text=black] {\large $\mathcal{H}_{1}$};$\draw[draw=black,solid,line width=2pt] (10.8, 1.2) -- (10.8, 0.3999999999999999);$$\draw[draw=black,solid,line width=2pt] (10.8, 1.2) -- (10.8, 2.0);$$\filldraw[draw=black,solid,line width=2pt,fill=myred!90] (10.8, 1.2) circle (0.4) node {};$\draw (10.8, 1.2) node[text=black] {\large $\mathcal{H}_{2}$};$\draw[draw=black,solid,line width=2pt] (12.0, 1.2) -- (12.0, 0.3999999999999999);$$\draw[draw=black,solid,line width=2pt] (12.0, 1.2) -- (12.0, 2.0);$$\filldraw[draw=black,solid,line width=2pt,fill=myred!90] (12.0, 1.2) circle (0.4) node {};$\draw (12.0, 1.2) node[text=black] {\large $\mathcal{H}_{3}$};$\draw[draw=black,solid,line width=2pt] (9.6, 2.4) -- (9.6, 1.5999999999999999);$$\draw[draw=black,solid,line width=2pt] (9.6, 2.4) -- (9.6, 3.2);$$\filldraw[draw=black,solid,line width=2pt,fill=myblue!90] (9.6, 2.4) circle (0.4) node {};$\draw (9.6, 2.4) node[text=black] {\large $\mathcal{X}_{1}$};$\draw[draw=black,solid,line width=2pt] (10.8, 2.4) -- (10.8, 1.5999999999999999);$$\draw[draw=black,solid,line width=2pt] (10.8, 2.4) -- (10.8, 3.2);$$\filldraw[draw=black,solid,line width=2pt,fill=myblue!90] (10.8, 2.4) circle (0.4) node {};$\draw (10.8, 2.4) node[text=black] {\large $\mathcal{X}_{2}$};$\draw[draw=black,solid,line width=2pt] (12.0, 2.4) -- (12.0, 1.5999999999999999);$$\draw[draw=black,solid,line width=2pt] (12.0, 2.4) -- (12.0, 3.2);$$\filldraw[draw=black,solid,line width=2pt,fill=myblue!90] (12.0, 2.4) circle (0.4) node {};$\draw (12.0, 2.4) node[text=black] {\large $\mathcal{X}_{3}$};$\draw[draw=black,solid,line width=2pt] (9.6, 3.5999999999999996) -- (9.6, 2.8);$$\draw[draw=black,solid,line width=2pt] (9.6, 3.5999999999999996) -- (9.6, 4.3999999999999995);$$\filldraw[draw=black,solid,line width=2pt,fill=myred!90] (9.6, 3.5999999999999996) circle (0.4) node {};$\draw (9.6, 3.5999999999999996) node[text=black] {\large $\mathcal{H}_{1}$};$\draw[draw=black,solid,line width=2pt] (10.8, 3.5999999999999996) -- (10.8, 2.8);$$\draw[draw=black,solid,line width=2pt] (10.8, 3.5999999999999996) -- (10.8, 4.3999999999999995);$$\filldraw[draw=black,solid,line width=2pt,fill=myred!90] (10.8, 3.5999999999999996) circle (0.4) node {};$\draw (10.8, 3.5999999999999996) node[text=black] {\large $\mathcal{H}_{2}$};$\draw[draw=black,solid,line width=2pt] (12.0, 3.5999999999999996) -- (12.0, 2.8);$$\draw[draw=black,solid,line width=2pt] (12.0, 3.5999999999999996) -- (12.0, 4.3999999999999995);$$\filldraw[draw=black,solid,line width=2pt,fill=myred!90] (12.0, 3.5999999999999996) circle (0.4) node {};$\draw (12.0, 3.5999999999999996) node[text=black] {\large $\mathcal{H}_{3}$};\draw (12.9, 2.4) node[text=black] {\large $=$};$\draw[draw=black,solid,line width=2pt] (13.799999999999999, 2.4) -- (15.0, 2.4);$$\draw[draw=black,solid,line width=2pt] (15.0, 2.4) -- (16.2, 2.4);$$\draw[draw=black,solid,line width=2pt] (13.799999999999999, 2.4) -- (13.799999999999999, 1.5999999999999999);$$\draw[draw=black,solid,line width=2pt] (13.799999999999999, 2.4) -- (13.799999999999999, 3.2);$$\filldraw[draw=black,solid,line width=2pt,fill=myyellow!90] (13.799999999999999, 2.4) circle (0.4) node {};$\draw (13.799999999999999, 2.4) node[text=black] {\large $b_{1}$};$\draw[draw=black,solid,line width=2pt] (15.0, 2.4) -- (15.0, 1.5999999999999999);$$\draw[draw=black,solid,line width=2pt] (15.0, 2.4) -- (15.0, 3.2);$$\filldraw[draw=black,solid,line width=2pt,fill=myyellow!90] (15.0, 2.4) circle (0.4) node {};$\draw (15.0, 2.4) node[text=black] {\large $b_{2}$};$\draw[draw=black,solid,line width=2pt] (16.2, 2.4) -- (16.2, 1.5999999999999999);$$\draw[draw=black,solid,line width=2pt] (16.2, 2.4) -- (16.2, 3.2);$$\filldraw[draw=black,solid,line width=2pt,fill=myyellow!90] (16.2, 2.4) circle (0.4) node {};$\draw (16.2, 2.4) node[text=black] {\large $b_{3}$};
    \end{tikzpicture}%
}
\caption[]{Tensor network representation of Eq. (\ref{eqn:linear_equation}) for solving for an approximate representation of the AGP as an MPO. For simplicity, we show an example with 3 sites, but the method generalizes to any number of sites (even an infinite number of sites). The tensors $b_i$ represent the right hand side term of Eq. (\ref{eqn:linear_equation}). This diagram represents the form of the linear equation that we can solve to obtain an approximation to the AGP parametrized by $\Xc_i$ (given $\Hc_i$ and $b_i$) using standard tensor network methods.}
\label{fig:linear_equation}
\end{figure*}
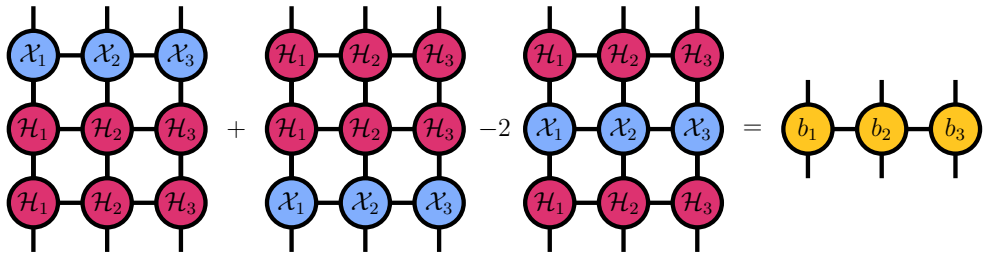

To compute the approximate AGP as an MPO, we first express $\Hc$ and $\partial_\l \Hc$ as MPOs. These MPOs can be efficiently constructed with low bond dimension (that is either constant or only grows polynomially with system size) for Hamiltonians with quasi-local interactions \cite{Pirvu_2010,chan2016matrix,hubig2017generic}. Then, we solve the linear equation in Eq. (\ref{eqn:linear_equation}) in terms of tensor networks (see Fig. \ref{fig:linear_equation}) to obtain $\Xc$ that approximates $\Ac_\l$.

We recast the linear equation to the form $Ax = b$, where $A$ is an MPO and $x, \, b$ are MPSs acting in a doubled Hilbert space. We convert $\Xc$ and the right side of Eq. (\ref{eqn:linear_equation}) to MPSs $\ket{\Xc}$ and $\ket{b}$, respectively, and recast the operators acting on $\Xc$ on the left side of the equation as a super-operator acting on $\ket{\Xc}$:
\begin{equation}\label{eqn:agp-linsolve}
	(I \otimes \Hc^2 + \Hc^2 \otimes I - 2\Hc \otimes \Hc) \ket{\Xc} = \ket{b} \to A \ket{\Xc} = \ket{b}\, ,
\end{equation}
where $A$ is an MPO acting on the doubled Hilbert space. We numerically solve this equation by using $\Ac_\l^{(1)} = i \alpha_1 \comm{\Hc}{\partial_\l \Hc}$ from Eq. (\ref{eqn:agp-comm}) as an initial guess to obtain the MPS $\ket{\Xc} \approx \ket{\Ac_\l}$ that approximately satisfies Eq. (\ref{eqn:agp-linsolve}). Specifically, we use a modification of the DMRG algorithm for solving linear equations of MPS \cite{jeckelmann2002dynamical,dolgov2014alternating} in order to approximately solve Eq. (\ref{eqn:agp-linsolve}). Finally, we recast $\ket{\Ac_\l}$ back as an MPO. The details of this method have been summarized in Alg. \ref{alg:agp}.

\begin{algorithm}[H]
\caption{Computing the AGP}\label{alg:agp}
\begin{algorithmic}
\Ensure $\Hc(\l) \, , \, \partial \Hc(\l)$
\State $A \gets I \otimes \Hc^2 + \Hc^2 \otimes I - 2 \Hc \otimes \Hc$
\State $\ket{b} \gets i\ket{\partial_\l \Hc \, \Hc - \Hc \,  \partial_\l \Hc}$
\State $\ket{\Xc_0} \gets i \alpha_1 \ket{\comm{\Hc}{\partial_\l \Hc}}$ \Comment{Use $\ell = 1$ in Eq. (\ref{eqn:agp-comm})}
\State $\ket{\Ac_\l} \gets \text{Solve for } \ket{\Xc} \text{ in } A \ket{\Xc}=\ket{b}$ \Comment{Initialize with $\ket{\Xc_0}$}
\State $\Ac_\l \gets \ket{\Ac_\l}$ \Comment{Recast MPS as an MPO}
\end{algorithmic}
\end{algorithm}

\begin{figure}[!htbp]
\centering
\includegraphics[width=\columnwidth]{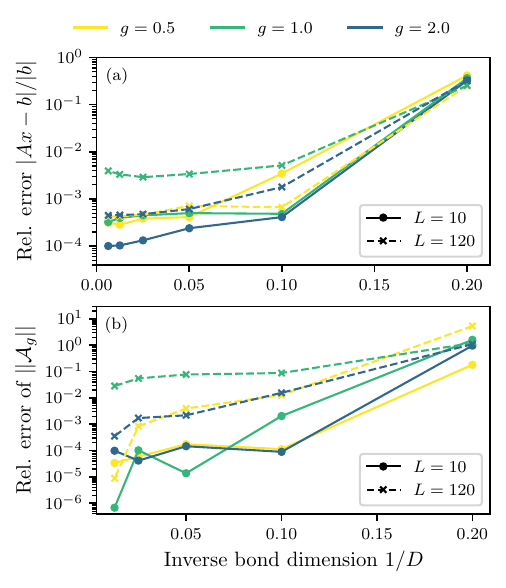}%
\caption[]{Convergence properties of solving for the MPO representation of the AGP with Alg. \ref{alg:agp}. We compute $\Ac_g$ as an MPO for $L = 10$ (shown in solid lines with dots) and $L = 120$ (shown in dashed lines with crosses). (a) shows the relative error $\abs{Ax-b}/\abs{b}$ determined by Eq. (\ref{eqn:agp-linsolve}) for various bond dimensions $D$ of $\Ac_g$. (b) shows the relative error of the norm of the MPO approximation of the AGP $\norm{\Ac_g}$ at bond dimension $D$ compared to the MPO approximation of $\norm{\Ac_g}$ computed at $D = 160$.}
\label{fig:tfim-agp}
\end{figure}

Now, we numerically construct the AGP as an MPO and study its properties as we vary the bond dimension $D$ of the AGP. To test the properties of Alg. \ref{alg:agp}, we consider the transverse field Ising model with open boundary conditions
\begin{align} \label{eqn:TFIM}
    \Hc_{\text{TFIM}} &= \sum^{L-1}_{i=1} \s^x_i \s^x_{i+1}  + g \sum^{L}_{i=1} \s^z_i \, , \\
    \partial_g \Hc_{\text{TFIM}} &= \sum^{L}_{i=1} \s^z_i \, ,
\end{align}
where $g$ is the transverse field strength and $L$ is the system size. Notably, a phase transition occurs at $g = 1$ ($g \leq 1$ when $L$ is finite): the eigenstates transition from a paramagnetic phase when $g > 1$ to an anti-ferromagnetic phase when $g < 1$. In Fig. \ref{fig:tfim-agp}, we construct $\Ac_g$ as an MPO using Alg. \ref{alg:agp} for $L = 10, 120$ at $g = 0.5, 1.0, 2.0$ with various bond dimensions $D = 5, 10, 20, 40, 80, 160$ of $\Ac_g$. As shown in Fig. \ref{fig:tfim-agp}a, the relative error $\abs{Ax-b}/\abs{b}$ generally decreases as $D$ increases for different realizations of $g$ and $L$ used. Furthermore, Fig. \ref{fig:tfim-agp}b shows similar convergence behaviors in $\norm{\Ac_g}$ as $D$ increases, even though we do not directly solve to minimize the relative error of $\norm{\Ac_g}$. Note that even near the critical point, where the bond dimension of the AGP would formally diverge in the thermodynamic limit, we can efficiently find approximate AGP with reasonable bond dimensions for this model. For example, on a personal computer, computing the AGP with bond dimension $D = 40$ at $L=120$ took a couple of minutes.

\section{\label{agp-evo}Adiabatic evolution of tensor networks with the AGP}

In this section, we briefly describe how adiabatic evolution of MPS can be performed efficiently using an MPO representation of the AGP, and describe its application to preparing states for iterative optimizations as well as scanning phase diagrams.

Following Alg. \ref{alg:agp}, we can construct the approximate AGP as an MPO for any quantum system with quasi-local Hamiltonian. In principle, we can use this AGP to adiabatically evolve any quantum state represented as an MPS to any parameter regime of the system. For example, as shown schematically in Fig. \ref{fig:propagate}, we can evolve MPS excitations from $\Hc(\l_i)$ to $\Hc(\l_f)$ by iteratively applying the AGPs on the states throughout the trajectory. Specifically, we use the time dependent variational principle method (TDVP) \cite{haegeman2011time,haegeman2016unifying} to adiabatically evolve the matrix product states, which can be formulated as a natural generalization of DMRG to performing evolution of an MPS and is well suited for evolving MPS with evolution generated by MPOs, unlike alternative MPS evolution methods like the time evolving block decimation (TEBD) \cite{vidal2004efficient} or time dependent DMRG \cite{white2004real} methods. In practice, we find that TDVP works very reliably for evolving MPS eigenstates with the AGP to perform adiabatic evolution. 

Now, we present two applications of using the AGP to perform adiabatic evolution of tensor networks. Firstly, we show how it can be used to improve the calculation of low energy states with DMRG by combining AGP evolution with DMRG. We also describe how AGP evolution of MPS can be used to automatically scan phase diagrams of low dimensional quantum systems.

Applying the technique introduced in Sec. \ref{agp-evo}, we combine our method for computing the MPO representation of the AGP (summarized in Alg. \ref{alg:agp}) with excited state DMRG with the goal of improving the computation time, reliability, and accuracy. Given that finding excited states (and even ground states) with DMRG can be computationally challenging around the critical point, where the states become nearly degenerate, we instead compute low energy MPS eigenstates sufficiently far away from the critical point, for example in a gapped phase where the states are less entangled and easier to find with DMRG or at a point in the phase diagram where a tensor network state can be efficiently constructed from another ansatz, such as a free fermion state \cite{kraus2010fermionic,evenbly2010entanglement,fishman2015compression,schuch2019matrix,wu2020tensor,petrica2021finite}, and adiabatically evolve these states towards the critical point.

To be specific, suppose that the critical point is located at $\l_f$ in Fig. \ref{fig:propagate}. We start by computing the low energy MPS at $\l_i$ with DMRG, where we choose $\l_i$ to be a point in the phase diagram far from the critical point where DMRG is reliable and fast. Then, we compute $\Ac_{\l_i}$ as an MPO using Alg. \ref{alg:agp} and evolve the states to $\l_i + \delta \l$ using TDVP. We compute a new AGP MPO at $\l_i + \delta \l$ (perhaps using the AGP we previously found at $\l_i$ as an initial guess for the optimization), and continue this procedure until we reach $\l_f$, at which point we perform excited state DMRG using the propagated MPS excitations as an ansatz to find excited states at $\l_f$. This serves as a controlled method for preparing good initial states for DMRG that are already close to the true low energy eigenstates, and therefore DMRG requires fewer iterations to converge and more reliably converges to states close to the true eigenstates. It is clear that the computational advantage of this method relies on the ability to (1) quickly find MPO approximations of AGPs, (2) efficiently evolve states using the MPO AGPs (for which we employ the TDVP method), and (3) maintaining accuracy of states during the evolution. As shown later in Sec. \ref{result}, we find that (1) and (2) can be easily achieved while (3) can be done by incorporating inexpensive excited state DMRG in the intermediate steps. The steps of this method combining DMRG with adiabatic evolution with the AGP are summarized in Alg. \ref{fig:alg}.

\begin{algorithm}[H]
\caption{AGP evolution of MPS}\label{fig:alg}
\begin{algorithmic}
\Require $\lambda_f > \lambda_i$
\State $\{\epsilon_n(\l_i),\ket{n(\l_i)}\} \gets $ DMRG
\State $\Ac_{\l_i} \gets \text{AGP}(\Ac_{\l_i}^{(1)}, \Hc(\lambda_i), \partial \Hc(\lambda_i))$ \Comment{Refer to Alg. \ref{alg:agp}}
\State $\lambda \gets \lambda_i$
\While{$\lambda < \lambda_f$}
\State $\delta \l \gets \delta \l \propto 1/\norm{\Ac_\l}$  
\State $\{\epsilon_n(\l),\ket{n(\l)}\} \gets \{\epsilon_n(\l + \delta \l),\ket{n(\l + \delta \l)}\} $ using TDVP
\State Optional: $\{\epsilon_n(\l),\ket{n(\l)} \} \gets $ DMRG on $\{\epsilon_n(\l),\ket{n(\l)} \}$
\State $\Ac_\l \gets \text{AGP}(\Ac_\l, \Hc(\lambda + \delta \l), \partial \Hc(\lambda + \delta \l))$
\State $\l \gets \l + \delta \l$
\EndWhile
\end{algorithmic}
\end{algorithm}

\begin{figure}[!htbp]
\centering
\includegraphics[width=\columnwidth]{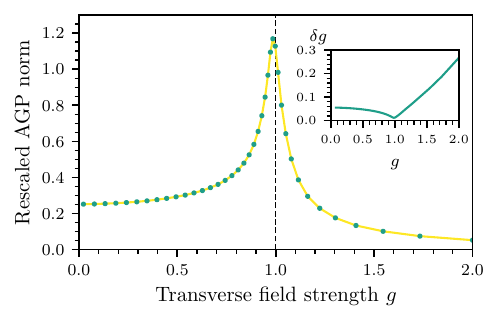}%
\caption[]{Norm of the AGP for the transverse field Ising model at system size $L = 120$. We scaled the norm of the AGP ($\norm{\Ac_g}$) by $1/\sqrt{L 2^L}$ to remove extensiveness due to $\partial_g \Hc$. The vertical line at $g = 1$ signifies the critical point and the inset shows the spacing $\delta g$ against $g$.}
\label{fig:agp-norm}
\end{figure}

Notably, the AGP calculations can be performed independently from the DMRG and TDVP computations given that the AGP evolution step sizes ($\delta \l$) can be determined by examining $\norm{\Ac_\l}$: as shown in Eqs. (\ref{eqn:agp}) and (\ref{eqn:agp-propagate}), the eigenstates change the most via adiabatic evolution when $\norm{\Ac_\l}$ is the largest. Interestingly, this usually coincides with the critical point as the norm of the AGP is equivalent to the fidelity susceptibility, which has been extensively used to probe quantum phase transitions and peaks or even diverges at the critical point \cite{venuti2007quantum,kolodrubetz2013classifying,zanardi2006ground,sierant2019fidelity}. Hence, we use smaller step sizes near the critical point by varying $\delta \l$ to be inversely proportional to $\norm{\Ac_\l}$ as shown in Alg. \ref{fig:alg} with an illustrative example shown in Fig. \ref{fig:agp-norm}. 

A modification of the method described above is to perform DMRG calculations at each intermediate value of $\l$'s between $\l_i$ and $\l_f$ and thereby compute accurate low-lying MPS excitations at each parameter point, where neither $\l_i$ nor $\l_f$ has to be at a the critical point. By varying $\delta \l$ to be inversely proportional to $\norm{\Ac_\l}$, the algorithm can naturally focus on potential regimes of criticality and so better understand the low energy dynamics of the system. Therefore, we can efficiently scan phase diagrams between any parameter regimes, even while crossing phase transitions or critical points. Additionally, we believe this method for adiabatic evolution of tensor networks with the AGP can find many applications beyond the ones described here, some of which we discuss in the conclusion.

\section{\label{result}Applications of AGP evolution of tensor networks}

\begin{figure*}[!htbp]
\centering
\includegraphics[width=2\columnwidth]{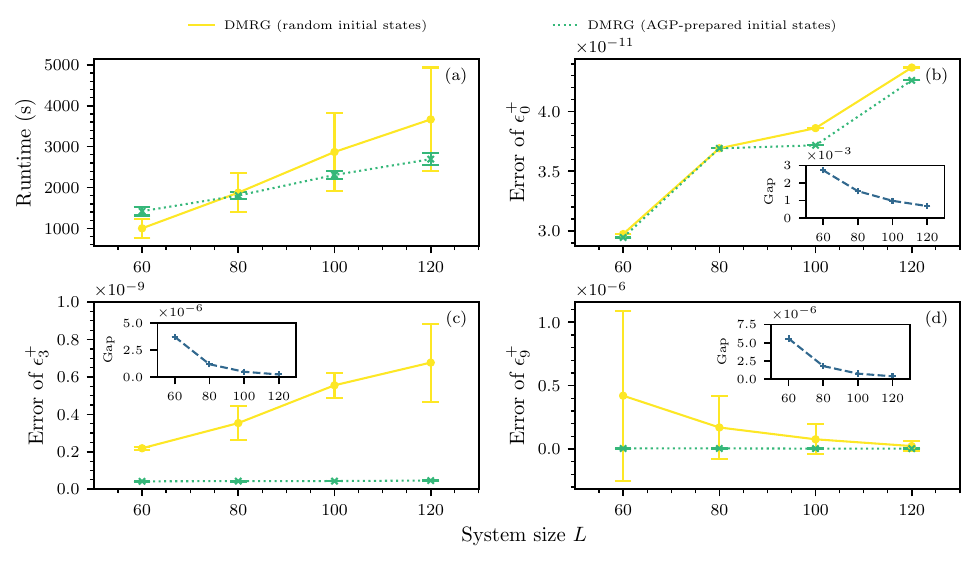}%
\caption[]{
Transverse field Ising model with $L = 120$ at transverse field strength $g = 1$. We compare the performance of DMRG with AGP-prepared initial states (dotted lines) against DMRG with random initial states (solid lines) at $g = 1$ for varying system sizes $L = 60,80,100,120$. Both methods find the ten lowest-lying states in even and odd sectors at $g=1$, respectively, until the relative energy difference between each sweep is less than $10^{-11}$. AGP-initialized DMRG used $8$ realizations of $g$, computed at $g=[2.0, 1.8, 1.6, 1.4, 1.2, 1.08, 1.02, 1.0]$. For the DMRG calculations, SVD truncation cutoffs between $10^{-11}$ and $10^{-8}$ are used for the MPS truncation while truncation cutoffs of $10^{-6}$ to $10^{-5}$ are used when solving for the AGP $\Ac_\l$. (a) shows the average runtime in seconds against system size. For AGP-initialized DMRG, this includes the time taken to obtain the AGP, perform the AGP evolutions, and execute the intermediary DMRG calculations. (b), (c), and (d) show the errors of $\epsilon_0^{+}$, $\epsilon_3^{+}$, and $\epsilon_9^{+}$, respectively. Insets in (b), (c), and (d) show the relative exact energy differences between $\epsilon_0^+$ and $\epsilon_1^{+}$, $\epsilon_3^+$ and $\epsilon_4^{+}$, and $\epsilon_8^+$ and $\epsilon_9^{+}$, respectively, for all system sizes considered.}
\label{fig:TFIM-benchmark}
\end{figure*}

In Sec. \ref{dmrg_agp_benchmark}, we demonstrate the utility of this approach for improving the reliability and performance of optimizing sets of low energy eigenstates with DMRG, focusing on the transverse field Ising model so that we can carefully benchmark our results. Then, in Sec. \ref{energyspectra}, we use AGP evolution of tensor networks, in combination with DMRG, to scan the phase diagram of the transverse field Ising model, with or without an integrability breaking longitudinal field, by adiabatically evolving a set of approximate low energy MPS eigenstates across the phase transition.

\subsection{\label{dmrg_agp_benchmark} Improving tensor network optimization with AGP evolution}

\begin{figure}[!htbp]
\centering
\includegraphics[width=\columnwidth]{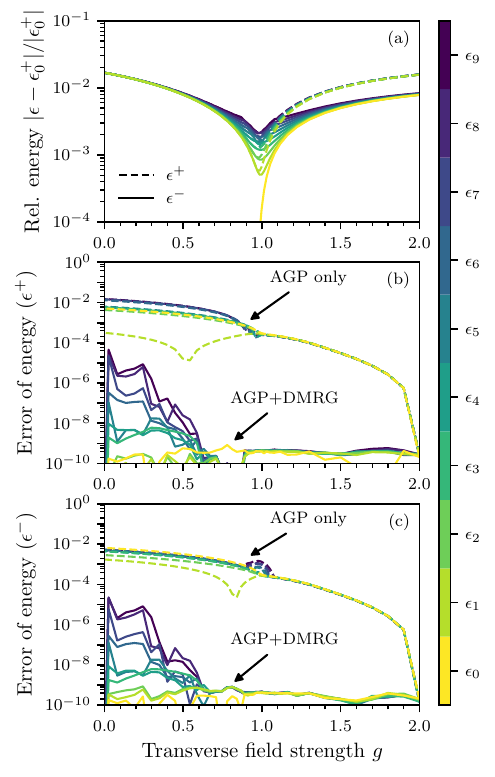}%
\caption[]{Transverse field Ising model at $L = 120$. We present the accuracy of our method by comparing the ten lowest-lying states (refer to color bar for ordering) from transverse field strength $g = 2$ to $g=0$, while passing the critical point at $g = 1$. We do this separately for even and odd parity sectors. In (a), we plot the relative energies in the even and odd sectors (denoted by dashed and solid lines, respectively) with respect to the ground state of the even sector ($\epsilon_0^{+}$) against $g$ obtained by using AGP+DMRG. In (b) and (c), we plot the error of the energy as a function of transverse field $g$, for even and odd sectors, respectively, where the dashed and solid lines represent data from using AGP evolution only and AGP+DMRG, respectively.}
\label{fig:TFIM-comparison}
\end{figure}

In this section, we demonstrate the utility of adiabatically evolving tensor networks with the AGP for improving the optimization of tensor networks. Many tensor network optimization methods, like DMRG, are iterative methods. The number of iterations needed to reach convergence, and whether or not the method converges to the global optimum, can depend on the initial state of the optimization. The convergence can be particularly sensitive to the initialization when there are other states very close to the globally optimal state, for example at or around quantum critical points, in which case the optimization can get stuck in a nearby state (i.e. a local minimum). Adiabatic evolution can provide a controlled way to prepare high quality initial states for iterative methods like DMRG. For example, if one can find high quality low energy states in a phase of the model that is relatively easy (for example, deep in a gapped phase, where the states generally have lower entanglement and the optimization is more reliable), one can then adiabatically evolve those states to a parameter regime that is more challenging and use them as initial states for an iterative optimization algorithm like DMRG.

Hence, we will focus on the problem of computing a set of low lying energy eigenstates near a critical point. For simplicity, we will focus on a particular flavor of excited state DMRG that finds higher excited states by adding weighted projectors of previously found states. One can then apply DMRG on the modified Hamiltonian of the form
\begin{equation}
    \Hc' = \Hc + w \sum_{m < n} \dyad{m}{m} \, ,
\end{equation}
where $w > 0$ is the penalty weight and the $n$th state is the desired excited state \cite{mcculloch2007density}. We emphasize that there are many other excited state DMRG techniques \cite{white1993density,ostlund1995thermodynamic,haegeman2012variational,banuls2013mass,hu2015excited,khemani2016obtaining,roberts2017implementation,van2021efficient,li2023tangent,li2023accurate} and, in principle, incorporating adiabatic evolution via the AGP to any iterative ground or excited state DMRG method to prepare initial states could improve the convergence, including the computational time and/or precision. 

We demonstrate the utility of AGP evolution for computing a set of low-lying excited states of the transverse field Ising model (Eq. (\ref{eqn:TFIM})) at the critical point, $g =1$. This model preserves $\s^z$ parity, which separates the Hamiltonian into even and odd parity sectors. In the following, we will denote the energy eigenvalues in the even sector as $\epsilon^+$ and in the odd sector as $\epsilon^{-}$. This model is integrable and is numerically solvable by a mapping to a free fermion model via the Jordan-Wigner transformation \cite{essler2016quench, mbeng2020quantum}, allowing us to exactly compute errors in the energies of MPSs.

For our approach that combines adiabatic evolution using the AGP and DMRG optimization, we start by running DMRG to find the ten lowest-lying states in even and odd sectors, respectively, at $g = 2$, until the relative energy difference between DMRG sweeps is less than $10^{-5}$. Then, we propagate those approximate eigenstates to $g=1$ using AGP evolution and DMRG calculations with only a few sweeps after each step adiabatic evolution. To ensure these intermediate DMRG runs do not dominate the computation time, they are run until the relative energy difference between sweeps is less than $10^{-6}$, which we find is enough to prevent accumulating large errors from taking large steps for the adiabatic evolution. Finally, we perform DMRG at $g = 1$ using the propagated states until the relative energy difference between sweeps for each eigenstate is less than $10^{-11}$, which is less than the relative difference between the exact energies of the eigenstates for the system sizes considered (see insets of Fig. \ref{fig:TFIM-benchmark}). This ensures that we have enough precision to properly find excited states at the critical point. In all, this procedure defines the AGP-initialized DMRG method.

For comparison, we run standard DMRG optimization at transverse field $g = 1$ using random initial wavefunctions. For this comparison, we used MPSs of bond dimension $10$ that are generated with random circuits, which we find is a relatively good unbiased method for initializing the starting states of the optimization. Then we perform DMRG to compute the ten lowest-lying states in even and odd sectors, respectively, until the relative energy difference between sweeps for each eigenstate is less than $10^{-11}$.

For DMRG with AGP-prepared initial states, we start with ten different realizations of random trial states at $g=2$. On the other hand, for DMRG with random initial states, we simply use twenty different realizations of random initial states at $g = 1$. Fig. \ref{fig:TFIM-benchmark}a shows the average runtimes of DMRG using random initial states compared to DMRG with initial states prepared using AGP evolution (with their sample standard deviations represented by error bars) for system sizes $L=60,80,100,$ and $120$. As shown, on average, AGP-initialized DMRG performs faster than the random-initialized DMRG for larger system sizes, as the runtime of the former increases slower with system size. Notably, the standard deviation of runtimes for standard DMRG drastically increases with system size, which further suggests that the AGP-initialized DMRG is a more reliable method. Next, we systematically compare the errors of the energies $\epsilon_0^{+}$, $\epsilon_3^{+}$, and $\epsilon_9^{+}$ in Figs. \ref{fig:TFIM-benchmark}b, \ref{fig:TFIM-benchmark}c, and \ref{fig:TFIM-benchmark}d, respectively, obtained using the two methods. As shown, on average, AGP-initialized DMRG outperforms random-initialized DMRG by achieving up to two orders of magnitude lower errors with much smaller fluctuations. As reference, the insets of Figs. \ref{fig:TFIM-benchmark}b, \ref{fig:TFIM-benchmark}c, and \ref{fig:TFIM-benchmark}d show the relative exact energy differences $\abs{\epsilon^+_0-\epsilon^+_1}/\abs{\epsilon^+_0}$, $\abs{\epsilon^+_3-\epsilon^+_4}/\abs{\epsilon^+_3}$, and $\abs{\epsilon^+_9-\epsilon^+_8}/\abs{\epsilon^+_9}$, respectively, for varying system sizes.  While not clearly shown, DMRG initialized with AGP-prepared states outperforms DMRG initialized with random states at $L = 120$ for the highest excited state studied in the even sector, where the errors are $2.2(16) \times 10^{-9}$ and $2.3(38) \times 10^{-8}$, respectively.

\subsection{\label{energyspectra}Scanning phase diagrams with AGP evolution of tensor networks}

\begin{figure*}[!htbp]
\centering
\includegraphics[width=2\columnwidth]
{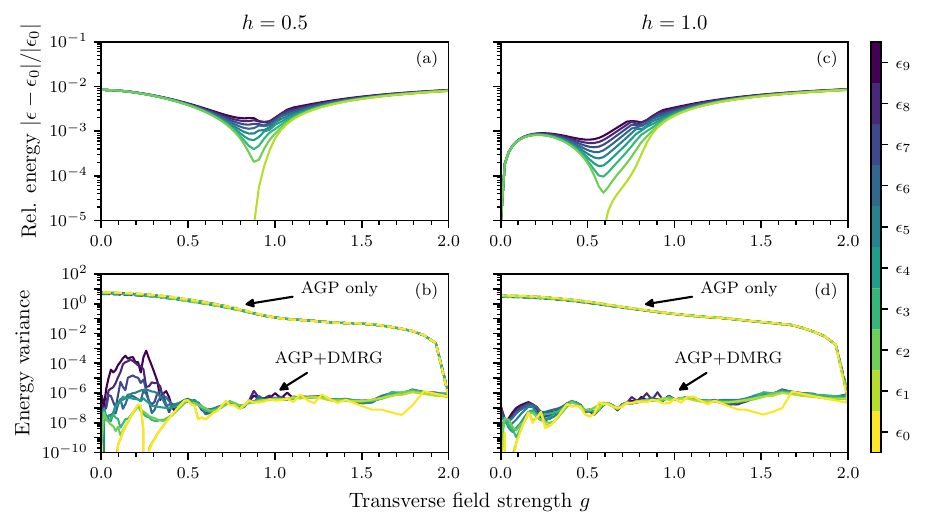}%
\caption[]{Ising model with transverse and longitudinal fields at $L = 120$ with longitudinal field strengths $h = 0.5, 1$. We compute the ten lowest-lying states (refer to color bar for ordering) from transverse field strength $g = 2$ to $g = 0$ at $h = 0.5$ in (a) and (b) and $h = 1$ in (c) and (d). (a) and (c) show the relative energy against the ground state energy $\epsilon_0$ for $h = 0.5$ and $h = 1$, respectively. (b) and (d) show the energy variance of the states ($\sigma_\epsilon^2 = \expval*{\Hc^2} - \expval{\Hc}^2$) for $h = 0.5$ and $h = 1.0$, respectively, where the solid lines represent data from using AGP+DMRG while dashed lines represent AGP evolution only.}
\label{fig:LTFIM-h_0dot5_1dot0}
\end{figure*}

Understanding the landscape of the system's phases, including the locations of quantum critical points, can provide critical information on numerous physical phenomena such as high-temperature superconductivity \cite{Dagotto1994correlated,Mapple1998high,orenstein2000advances}, quantum hall states \cite{Sondhi1997continuous,piazza1999first}, magnetic insulators \cite{Bitko1996quantum}, and much more. However, finding low-lying energy eigenstates (and thereby studying the phases) becomes particularly challenging near phase transitions, where the correlation length generally diverges. Hence, scanning phase diagrams is generally difficult if passing through different phases. While standard DMRG procedure can be employed, DMRG optimization can become expensive and difficult to converge near phase transitions. To prevent this, we augment our AGP evolution procedure by adding accurate DMRG calculations along the trajectory (coined AGP+DMRG) to find low-lying energy eigenstates. This makes finding phase diagrams more computationally efficient and robust, for using AGP-initialized states as ansatzes is more reliable than random trial states. Finally, using the AGP provides automated scanning of the phase diagram by automatically using more steps near the phase transitions.

Using AGP+DMRG, we compute the low-lying energy spectra of the transverse field Ising model and propagate them to scan its phase diagrams. In the transverse field Ising model, we pass through the critical point at $g = 1$, starting at $g = 2$ and terminating at $g = 0$, while propagating ten lowest-lying states in even and odd sectors, respectively, with system size $L = 120$. Specifically, $40$ values of $g$'s are considered with similar spacing as shown in Fig. \ref{fig:agp-norm} but with $\delta g$ capped at $0.1$ instead (we find that allowing larger steps than that leads to large errors in the adiabatic evolution). In contrast to Sec. \ref{dmrg_agp_benchmark}, however, we perform more expensive intermediary DMRG calculations to ensure more accurate results for all values of $g$ considered. The relative energies with respect to the ground state energy ($\epsilon_0^{+}$) are shown in Fig. \ref{fig:TFIM-comparison}. As shown, when $g \lesssim 1$, $\epsilon_i^{+} \approx \epsilon_i^{-}$ for all integer $i$'s such that $0 \leq i \leq 9$. Furthermore, as shown in Figs. \ref{fig:TFIM-comparison}b and \ref{fig:TFIM-comparison}c, incorporating intermediary DMRG calculations in between AGP evolutions drastically improves the results. Importantly, AGP+DMRG is able to correctly calculate the excited states even while passing through the critical point. While not shown, AGP calculation and evolution times only contributed to roughly $10\%$ of the total computation time with the remainder taken up by DMRG. On average, intermediary DMRG calculations at each value of $g$ took roughly $88\%$ less time than the DMRG calculations at starting point $g=2$. It is clear that AGP-related computation is hardly the bottleneck of  AGP+DMRG.

Now, we consider the non-integrable Ising model by adding a longitudinal field with strength $h$ to our Hamiltionian in Eq. (\ref{eqn:TFIM}):
\begin{equation}\label{eqn:LTFIM}
    \Hc_\text{LTFIM} = \sum^{L-1}_{i=1} \s^x_i \s^x_{i+1}  + g \sum^{L}_{i=1} \s^z_i + h \sum^{L}_{i=1} \s^x_i \, .
\end{equation}
Contrary to the previous model, this Hamiltonian is not exactly solvable, generically has no symmetry sectors (open boundary conditions are employed), and is generally chaotic \cite{kim2014testing}. This model has a more complicated phase structure and exhibits phase transitions at $0 \leq g \leq 1$ when $0 \leq h \leq 2$. Here, we employ AGP+DMRG to study the low-lying spectrum of $\Hc_\text{LTFIM}$ as we pass through phase transitions for system size $L = 120$ at two fixed values of $h=0.5,1$. In Fig. \ref{fig:LTFIM-h_0dot5_1dot0}, we propagate ten lowest-lying states from $g = 2$ to $g = 0$ at $h = 0.5$ and $h = 1$. In Figs. \ref{fig:LTFIM-h_0dot5_1dot0}a and \ref{fig:LTFIM-h_0dot5_1dot0}c, we plot the energy spectrum of the states relative to the ground state energy for $h =0.5$ and $h=1$, respectively. Notably, the energy spectrum shown in Fig. \ref{fig:LTFIM-h_0dot5_1dot0}a is similar to that shown in Fig. \ref{fig:TFIM-comparison}c, with slight differences in values of $g$ at which the ground state becomes doubly degenerate. Similar to the integrable case, the error increases when $g \lesssim 0.5$ as the excited states become highly degenerate. However, in Fig. \ref{fig:LTFIM-h_0dot5_1dot0}c, we have a noticeably different energy spectrum with all of the ten-lowest lying states becoming degenerate at $g = 0$. As we lack access to exact energies, we compute the energy variance $\sigma_\epsilon^2 = \expval*{\Hc^2} - \expval{\Hc}^2$ for $h =0.5$ and $h = 1$ in Figs. \ref{fig:LTFIM-h_0dot5_1dot0}b and \ref{fig:LTFIM-h_0dot5_1dot0}d, respectively, where the expectation values are taken with respect to computed states $\ket{n}$ such that $\Hc \ket{n} \approx \epsilon_n \ket{n}$. The variance provides a measure of how close $\ket{n}$ is to being an eigenstate of $\Hc$. As shown, a variance on the order of $10^{-6}$ is achievable using AGP+DMRG. Once again, the AGP calculation and evolution contributed to roughly $12\%$ of the total runtime with DMRG calculations taking up the rest. Further, on average, intermediary DMRG calculations at each value of $g$ took roughly $80\%$ less time than the DMRG calculations at $g = 2$.

\section{Conclusion}
We presented a tensor network method for computing the adiabatic gauge potential, the generator of adiabatic eigenstate deformations, as a tensor network and tested the method on some paradigmatic models. For simplicity, a matrix product operator (MPO) ansatz was used, though generalizing to other tensor networks (or other ansatzes) is straightforward.

Future work and applications are abundant. Firstly, our choice of the particular excited state DMRG method used is arbitrary and, in principle, adiabatically evolving matrix product states with the adiabatic gauge potential has the possibility to improve other excited state (or ground state) DMRG algorithms. Here are more applications of the AGP in tensor networks:
\begin{enumerate}
    \item Some recent excited state DMRG methods could directly benefit from our results such as for adiabatic transport of orbital wavefunctions in molecular crystals \cite{hu2015excited} and multi-target matrix product state ansatz \cite{li2023accurate}.
    \item The AGP could be applied to other variational tensor network methods: other excited state methods, 2D tensor network state (TNS) or projected entangled pair states (PEPS) methods, preparation of starting states for dynamics, and more.
    \item Continuum systems of weakly interacting electrons can be slow to converge with DMRG \cite{stoudenmire2012one,dolfi2012multigrid,miles2017sliced,haghshenas2021numerical,dutta2022density}. This convergence could be improved by preparing a free fermion state as a tensor network \cite{kraus2010fermionic,evenbly2010entanglement,fishman2015compression,schuch2019matrix,wu2020tensor,petrica2021finite}, evolving it with the AGP to an interacting regime, and then performing DMRG using the AGP-prepared state.
    \item State preparation for quantum computers could be implemented using the AGP. Namely, one could prepare an MPS or TNS, adiabatically evolve it to the desired parameter regime using the AGP, and then convert it to a low-depth circuit \cite{ran2020encoding,slattery2021quantum}. This could help produce starting states for simulations with quantum chemistry \cite{kassal2008polynomial,lanyon2010towards}, quantum field theory \cite{moosavian2018faster,preskill2018simulating}, variational quantum circuits \cite{huggins2019towards,liu2019variational,fedorov2022vqe,dborin2022matrix,xu2022differentiable,hou2023sequentially}, and more.
    \item A natural extension of our work lies on the tensor network simulations of adiabatic quantum algorithms: quantum approximate optimization algorithms (QAOAs) \cite{farhi2014quantum}, Grover's algorithm \cite{grover1996fast,stoudenmire2023grovers}, and adiabatic quantum computation \cite{farhi2000quantum}.
    \item The AGP as an MPO can be generalized to more Hamiltonian parameters and be used to study, for example, geodesics in parameter space \cite{sugiura2021adiabatic,kim2023integrability} for larger system sizes.
    \item In Ref. \cite{herbst2022surrogate}, they use a sampling of states in phase space, determined by computing the residuals of high-dimensional eigenvalue problems, to map out the rest of the phase diagram. Instead, the AGP could be used to automate this sampling process to lower the computational cost.
    \item Ref. \cite{yarloo2023adiabatic} showed that adiabatic time evolution of highly excited states, particularly quantum scars, can be robustly performed. Using this, our work could be extended to adiabatically evolving highly excited states with the AGP.
\end{enumerate}

\textit{Note added}: Few days after our submission to \texttt{arXiv}, another article \cite{keever2023adiabatic} appeared on \texttt{arXiv} that uses a similar construction to variationally prepare the AGP as an MPO. Complementary to our results, their work focuses on using the AGP to optimize quantum circuits for adiabatic quantum computing. 

\vspace{1mm}

\subsection*{Computing Resources and Software Packages}
The code used to produce the numerical results in this paper was written using the \textbf{ITensorTDVP.jl} package \cite{ITensorTDVP} --- a publicly available Julia \cite{bezanson2012julia} package for solving a variety of equations in terms of MPS including finding eigenstates with DMRG and DMRG-X \cite{Khemani2016}, performing time evolution with TDVP, and solving linear equations of MPS. It is built on top of \textbf{ITensors.jl} \cite{itensor-r0.3}, which provides the basic tensor operations and MPO/MPS types. For the MPS linear solver in ITensorTDVP.jl, we use the \emph{linsolve} method from \textbf{KrylovKit.jl} \cite{KrylovKit} as the local solver, which implements the GMRES algorithm \cite{GMRES}. Calculations of the AGPs for all of the numerical computations were performed on a personal laptop while the rest of the numerics were performed on Flatiron Institute's Rusty computing cluster. The code to construct the AGP as an MPO is available in \textbf{ITensorAGP.jl} \cite{ITensorAGP}.
\par The tensor network diagrams in this paper were produced using the publicly available package \textbf{GraphTikZ.jl} \cite{GraphTikz}, a general-purpose Julia package for visualizing graphs, including tensor networks.

\begin{acknowledgments}
The authors thank Miles Stoudenmire and Anatoli Polkovnikov for helpful discussions. M.F. and D.S. are grateful for ongoing support through the Flatiron Institute, a division of the Simons Foundation. H.K. acknowledges partial support from NSF (grant no. DMR-2103658). D.S. is partially supported by AFOSR (grant no. FA9550-21-1-0236) and NSF (grant no. OAC-2118310).
\end{acknowledgments}

\bibliography{bibliography.bib}

\end{document}